\documentstyle[aps,preprint,eqsecnum]{revtex} \baselineskip 12pt
\begin{document}

\title{LOCAL BRS TRANSFORMATIONS AND $OSp(3,1|2)$ SYMMETRY OF
YANG-MILLS THEORIES}

\vspace{.7in}

\author{ Satish D. Joglekar\footnote{e-mail address:- sdj@iitk.ernet.in}
and Bhabani Prasad Mandal\footnote{Address after 10th December:-
Institute of Physics, Schivalaya Marg, Bhubaneswar-751005, India}}

\vspace{.7in}

\address{Department of physics \\ Indian Institute of Technology, Kanpur\\
Kanpur, 208016, (INDIA)}

\maketitle

\vspace{.8in}

\begin{abstract}
We consider a certain local generalization of BRS transformations
of Yang-Mills theory in which the anti-commuting parameter is
space time dependent. While these are not exact symmetries, they do
lead to a new nontrivial WT identity. We make a precise connection
between the ``local BRS "and the broken orthosymplectic symmetry
recently found in superspace formulation of Yang-Mills theory by
showing that the local BRS WT identity is precisely the WT identity
obtained in the superspace formulation via a superrotation. This
``local BRS " WT identity could lead to new consequences not contained
in the usual BRS WT identity.
\end{abstract}
\newpage

\section{INTRODUCTION}

The theories with non-abelian local gauge invariance
have acquired a central place in particle physics as
they enter into the unified description strong, weak
and electromagnetic interactions viz. the standard
model. Discussion related to the gauge invariance
and its preservation to all orders are of a primary
importance as the unitarity and renormalizability of
all these theories depend crutically on these. The consequence
of the local gauge invariance are formulated as a series of
Ward- Takahashi (WT) identities and these enter largely in the
above discussion. While the initial formulations of the
consequences of the local invariance (the WT identities )
depend solely on the local gauge transformation of the gauge
field~\cite{le}, these formulation were later replaced by much more
elegent techniques using the BRS symmetry discovered later~\cite{zinn,lee}.
This
BRS symmetry is a global symmetry ( i.e. depends on a space time
{\it independent} anti-commuting parameter.) The relevent
consequences of the {\it local} gauge invariance of the
basic Yang-Mills action $ \frac{1}{4} \int d^4x F_{\mu\nu}
F^{\mu\nu }$ that are necessary for the discussion of
renormalization of gauge theories and their unitarity can also
be derived via latter {\it global} symmetry, the BRS symmetry~\cite{lee}. A
natural question that comes to mind is whether the
{\it global}
 BRS transformations yield {\it all} the information
 contained the original {\it local} gauge invariance of
 the theory.

 The question becomes relevant in view of a recent observation
 by the authors \cite{osp}. It was in connection with the
 derivationof WT identities in gauge theories in a superspace
 formulation. A superspace formulation of gauge theories has
 been constructed that has a (broken) $ OSp(3,1|2) $ invariance
 corresponding to rotations in a six dimensional superspace that
 mix $x^\mu $ and the anti-commuting coordinates $ \lambda $ and
 $ \theta $. In this formulation the BRS transformation of the
 original Yang-Mills theory been replaced in effect by a
 specific kind of superrotation. The broken  $ OSp(3,1|2) $ then
 leads to an equation of the form \footnote{This differs slightly from
 the WT identity in Ref.~\cite{osp} partly because $\bar{W}$ used is
 different.}
 \begin{equation}
 \bar{W} [\bar{K} ,t] = \bar{W} [\bar{K} ^\prime ,t^\prime ] -
 \ll \int \delta _4[\frac{ \partial }{ \partial \theta }( \partial
 ^\mu A_\mu \zeta)] d^4x\gg
 \label{ww}
 \end{equation}
 Then, as a {\it special} piece of information, it was
 shown from this that
 \begin{equation}
 \frac{ \partial \bar{W} }{ \partial \theta } =0
 \label{wt}
 \end{equation}
and this was shown to embody the WT identities of the gauge
theories. The question naturally arises if the Eq.~\ref{ww}
embodying broken $ OSp(3,1|2) $ symmetry contains a additional
information not contained in Eq.~\ref{wt} which are equivalent
to the usual BRS WT identities.

In searching for this question, we can do a ``backward''
derivation of the existance of an ``approximate'' $ OSp(3,1|2) $
invariance that can be associated with the Yang-Mills theory.
This derivation makes use of what we call ``local BRS'' and these
are transformation that are a local generalization of the usual
BRS transformation viz.
\begin{eqnarray}
\delta A_\mu (x) &=& D_\mu c(x)\,\, \Lambda (x) \nonumber \\
\delta c & = & -\frac{1}{2} gfcc \,\,\Lambda (x) \nonumber \\
\delta \zeta  &=& -\frac{1}{ \lambda } \partial\cdot A \,\,\Lambda (x)
\label{lb}
\end{eqnarray}
Here, $ \Lambda (x)$ is an x-dependent arbitary Grassmannian.
We hasten to add that the transformations of Eq.~\ref{lb} are
{\it not} a symmetry of the Yang-Mills effective action
$ S_{eff}$ [ nor is the first of Eq.~\ref{lb} a special case
of the local gauge transformation on $A_\mu$ .] However, we can
look upon transformation of Eq.~\ref{lb} as a ``broken'' or
approximate symmetry of $ S_{eff}$ and use it to obtain its
cosequences via the usual procedure for obtaining WT identities
. [Note that {\it any} field transformation on the
integration variables of the generating functional does give
some identity : whether it is useful or not depends on whether
the transformation is a ``relevant'' one for the action under
consideration.] The consequence sources obtained are seen in
particular to contain the origin of the broken $ OSp(3,1|2) $
symmetry found in superspace.

We should remark, however, that the additional results
contained in the local BRS WT identity, while they do explain
the origin of $ OSp(3,1|2) $ symmetry, do not in an easy way,
lend to consequences; as the identity contains new operators
which are not already included in the generating functional
. However, the new operator is a BRS variation of a simple
operator of dimension three and is expected to be
multiplicatively renormalizable. Further investigation is needed
to come up with the new additional results.

Now we briefly state the plan of the paper. In sec.II we introduce
the notations and the ``local BRS" transformations. We also introduce
briefly the superspace formulation of Ref. 4. In sec. III, we derive
the ``local BRS " WT identities. In sec. IV, we show the connection
between  these  and the OSp WT identity of Eq.~\ref{ww}. In sec. V,
we summerize the results and make a few obeservations.

\section{Preliminary}
\subsection{Notations}
In this section we shall introduce the notations and conventions
which will be used frequently in this paper. We consider the
generating functional for the Yang-Mills theory in linear gauges
as follows
\begin{eqnarray}
W &=& \int \left [ dA \, dc \, d \zeta \right ]\exp \left ( iS + i\int d^4x
\left\{ j_\mu ^\alpha A^\alpha _\mu +\bar{\xi}^\alpha c^\alpha +
\zeta ^\alpha \xi ^\alpha + \kappa ^{\alpha \mu} D^{\alpha \beta
}_\mu c^ \beta +\nonumber \right.\right.\\&& \left.\left.\mbox{\hspace{3.3in}}
\frac{g}{2} l ^\alpha f^{\alpha \beta \gamma }
c^ \beta c^ \gamma \right\}\right )\nonumber \\
&=& W\left [ j,\bar{\xi}, \xi , \kappa ,-l,t\right ]
\label{ww1}
\end{eqnarray}
Where the action, $S $ is given by
\begin{equation}
S = S_0 + S_g + S_{gf}
\label{s}
\end{equation}
with
\begin{eqnarray}
S_0 &=& \int d^4x\left\{ -\frac{1}{4} F^\alpha _{\mu\nu}F^{\alpha
\mu\nu }\right\}\nonumber \\
S_g &=& \int d^4x \left\{ - \partial ^\mu \zeta ^\alpha D^{\alpha
\beta }_\mu c^\alpha \right\} \nonumber \\
S_{gf} &=& \left\{ -\frac{1}{2 \lambda }( \partial .A^\alpha
+t^\alpha)^2\right \}
\end{eqnarray}
Here we assume a Yang-Mills theory with a simple gauge group
and introduce following notations
\begin{eqnarray}
\mbox{Lie Algebra :} \left[ T^\alpha ,\, T^ \beta \right] &=&
if^{\alpha \beta \gamma} T^\gamma \nonumber \\
\mbox{Covariant Derivative :} D^{\alpha \beta }_\mu c^ \beta &=&
\left( - \partial _\mu \delta ^{\alpha \beta}c^ \beta  +gf^{\alpha \beta
\gamma} A^\gamma _\mu c^ \beta \right )
\end{eqnarray}
$ f^{\alpha \beta \gamma }$ are structure constant of the gauge
group and totally anti symmetric.

\subsection{Local BRS Transformations}

Now we introduce the local BRS transformations which we shall
use in next section to derive WT identities.

The local BRS transformations are,
\begin{eqnarray}
\delta A^\alpha _\mu &=& D^{\alpha \beta }_\mu c^ \beta \Lambda
(x)\nonumber \\
&=& D^{\alpha \beta }_\mu \left( c^ \beta \Lambda (x)\right) +
c^\alpha \partial _\mu \Lambda (x) \nonumber \\
\delta c^\alpha  &=& -\frac{g}{2} f^{\alpha \beta \gamma }c^
\beta c^ \gamma \Lambda (x) \nonumber\\
\delta \zeta ^\alpha  &= &-\frac{1}{ \lambda }( \partial\cdot A^\alpha
+t^\alpha ) \Lambda (x)
\label{lbt}
\end{eqnarray}
where the anti-commuting BRS parameter, $\Lambda (x) $ is function
of x.

Note in the transformations ~\ref{lbt} spatial differential
operator is involved only in the transformation of gauge field
$ A^\alpha _\mu .$ We can perform these local BRS transformation
in two steps as follows

First consider the transformation, $ \delta _1$
\begin{eqnarray}
\delta_1 A^\alpha _\mu &=& D^{\alpha \beta }_\mu\left( c^ \beta \Lambda
(x)\right)\nonumber \\
\delta_1 c^\alpha  &=& -\frac{g}{2} f^{\alpha \beta \gamma }c^
\beta c^ \gamma \Lambda (x) \nonumber\\
\delta _1\zeta ^\alpha  &=& -\frac{1}{ \lambda }( \partial\cdot A^\alpha
+t^\alpha ) \Lambda (x)
\label{lb1}
\end{eqnarray}
and then consider the transformation $\delta _2$
\begin{eqnarray}
\delta _2 A^\alpha _\mu &=& c^\alpha \partial _\mu \Lambda
(x)\nonumber \\
\delta _2 c^\alpha &=& 0\nonumber \\
\delta _2 \zeta ^\alpha &=& 0
\label{lb2}
\end{eqnarray}
Where, $ \delta $ of~\ref{lbt} is simply $ \delta = \delta _1+
\delta _2 $. Now the transformation~\ref{lb1} on $ A_\mu$ is
still a gauge transformation and symmetry of $S_0$.
Hence the effect of local BRS transformation on $ S_0$ can be
obtained directly using transformation~\ref{lb2} only which
is very simple in nature.

Now we shall see how the action in ~\ref{s} transform under
these local BRS transformations. We only write down the changes
of the action here.

Under the transformations ~\ref{lbt}

\begin{eqnarray}
\delta S_0 = \delta _2 S_0 & = &\int d^4x \,\,c \partial _\mu \Lambda
(x) \frac{ \delta S_0}{ \delta A_\mu (x)}\mbox{\hspace{.4in}}
\nonumber \\
&=& \int d^4x \partial _\mu \left( D_\nu c F^{\mu\nu}\right) \Lambda
(x) \mbox{\hspace{1in}}\label{ni11}\\
\delta S_g + \delta S_{g.f} &=& \int d^4x\left\{ \partial _\mu
\left( \frac{ (\partial\cdot A+t)}{ \lambda } D_\mu c \right) \Lambda (x)
+ \partial _\mu \left( \partial ^\mu \zeta \frac{g}{2}fcc\right)
\Lambda (x) \right\}\nonumber \\
&\equiv & \delta _1\left( S_g +S_{g.f}\right) + \delta _2\left( S_g +
S_{g.f} \right)\label{ds} \\
\mbox{ with     } \nonumber &&\\
\delta _1\left( S_g +S_{g.f}\right) &= & \int d^4x \left\{ \partial _\mu
\left[ -\frac{g}{2} \partial^\mu \zeta fcc\right] \Lambda \right
.\nonumber  \\
&& \left.+\partial _\mu\left [\frac{( \partial \cdot A +t)}{ \lambda }D_\mu
 c \right ] \Lambda + \partial _\mu \left [ \partial _\mu \frac{(
 \partial\cdot A +t)}{ \lambda } c \right ] \Lambda
 \right\} \label{ni13}
\end{eqnarray}
One can check the relations in Eq.~\ref{ds} very easily. We
shall use these transformation properties in next section for
the derivation of WT identities.

\subsection{The Superspace Formulation }
We briefly recapitulate the superspace formulation of
Ref.~\cite{sdj}. In this superspace one has four space-time
commuting and two anti-commuting dimensions sources that
$\bar{x} ^i = ( x^\mu ,\lambda, \theta ). $ The metric in this
space is $ g_{ij} = g_{\mu\nu} = \mbox{diag} ( 1,-1,-1,-1)$ for
$ 0\leq i ,\,j \leq 3$ and $ g_{ij} = -\epsilon _{ij} \,\,(
4\leq i,\, j\leq 5$ with $ \epsilon _{45} = 1$. The group of
transformation that preserves $ \bar{x}^i \bar{x} _i$ is $
OSp(3,1|2) $. We introduce the superfields $ \bar{A}^\alpha _i
(\bar{x}) $ and $ \zeta ^\alpha (\bar{x}) $ transforming as a
covariant vector and a scalar under $ OSp(3,1|2) $ . The $
A^\alpha _\mu (x) $ are identified with the usual gauge fields,
 $ A^\alpha _5(x) \equiv c^\alpha _5(x) $ with the usual ghost fields
 , $ A^\alpha _4 (x)\equiv c_4 ^\alpha (x)$ is an additional
 ghost field and $ \zeta (x) $ is the usual antighost field of
 the gauge theories. The remaining fields in the expansion of
 $ \bar{A}^\alpha _i (\bar{x}) $ and $ \zeta ^\alpha (\bar{x}) $
 in terms of $ \lambda $ and $ \theta $ are certain auxiliary
 fields. We also introduce a vector and a scalar supersource
 $\bar{K} ^{\alpha i}(\bar{x}) $ and $ t^\alpha  (\bar{x}) $.
 As was shown in Ref.\cite{sdj}. $ \bar{K} ^{\alpha i} (\bar{x}) $
 contains in it compactly sources for gauge, ghost fields {\it
 and also } sources for composite fields $ \kappa _\mu $ and $l$
 all in a single supermultiplet of $ OSp(3,1|2) $. $t^\alpha
 (\bar{x}) $ contains in it the source for antighost field and
 the source for the gauge fixing $(\frac{1}{\sqrt{\eta_0}})
 \partial ^\mu A_\mu ^\alpha $ in one multiplet. An action for
 this superfield theory was introduced as
 \begin{eqnarray}
 \bar{S}&= &\int d^4x {\bar{{\cal L}}}_0[\bar{A}] + \int d^4x
 \frac{ \partial }{ \partial \theta }\left\{ \bar{K} ^{\alpha i}
 (\bar{x}) \bar{A}^\alpha _i (\bar{x}) + \zeta ^\alpha (\bar{x})
 \left[ \partial ^\mu A_\mu ^\alpha (\bar{x})  +\frac{1}{2\eta_0}
 \zeta _{, \theta }^\alpha (\bar{x}) +t^\alpha (\bar{x}) \right]\right\}
 \nonumber \\
 &\equiv  &\bar{S}_0 +\bar{S}_1
\label{sa}
\end{eqnarray}
with $ {\bar {\cal L}}_0 [\bar{A}] = -\frac{1}{4} g^{ik}g^{jl}
\bar{F}^\alpha _{ij} \bar{F}^\alpha _{kl} $ and $ \bar{F}^\alpha
_{ij} = \partial _i \bar{A}^\alpha _j - \bar{A}^\alpha _i
\partial _j + g_0 f^{\alpha \beta \gamma } \bar{A}^ \beta _i
\bar{A}^ \gamma _j $ and a generating functional was introduced
as ( please refer to \cite{sdj} for details of definition of
measure)
\begin{equation}
\bar{W} [ \bar{K} ,\bar{t}] = \int \{d\bar{A}\}\{ \bar{\zeta }
\}\exp{\left(iS[\bar{A}, \zeta \bar{K} ,t ]\right)}
\label{sw}
\end{equation}
In Ref. \cite{sdj}, this generating functional was shown to contain
the generating functional of Green's function of gauge theory,
viz.,
\begin{equation}
\bar{W} [\bar{K} ,t] =\prod _{\alpha ,y} \delta (K^{\alpha 4}(y))
\prod _{ \beta ,z} K^{ \beta 4}_{, \theta }(z) W[ K^{\alpha
\mu}_{, \theta };K^{\alpha 5}_{, \theta };-t^\alpha _{, \theta
}; K^{\alpha \mu}; K^{\alpha 5};t^\alpha ]
\label{t275}
\end{equation}
where the usual generating functional for gauge theories
is defined by Eq. \ref{w}
Thus $ K^{\alpha \mu}_{, \theta }; K^{\alpha 5}_{, \theta };
-t^\alpha _{, \theta } $ serving as sources for gauge, ghost and
antighost fields respectively and $ K^{\alpha \mu}, K^{\alpha
5},t^\alpha $  serving as sources for operators corresponding
to BRS variations, all are in compact supermultiplets.

We note some of the virtues of this formulation (not directly
used in this work) : (i)$ {\bar{{\cal L}}}_0[\bar{A}]$ is $
OSp(3,1|2) $ invariant. $ \bar{S}_0 $ is invariant under
infinitesimal $ OSp(3,1|2) $ . (ii) The terms $\bar{S}_1$ that
breaks $ OSp(3,1|2) $ invariance has a relatively simple form
and, moreover, has a partial $ OSp(3,1|2) $ invariance [Lorentz
transformations, a subgroup of Sp(2) and certain superspace
rotations] (iii) The measure is invariant under infinitesimal
$ OSp(3,1|2) $ and $ \bar{W} [\bar{K} ,\bar{t}] $ has the
invariances listed under (ii).
\section{WT Identities Under Local BRS Transformations}
Here in this section we derive the first result of this paper,
i.e., the local WT identities satisfied by the generating
functional, $ W $ using local BRS invariance.

Performing the ``local BRS'' transformation of Eq.~\ref{lbt} in
the generating functional of Eq.~\ref{ww1} for the Yang-MIlls
theory and equating the changes to zero, we obtained
\begin{equation}
\delta S +\int K^\mu \delta (D_\mu c)d^4x +\int d^4x\left
[ j^\mu \delta A^\mu +\bar{\xi} \delta c + \delta \zeta \xi\right]
= 0
\end{equation}
where we have used the fact that $ \delta (\frac{g}{2}f^{\alpha
\beta \gamma } c^ \beta c^ \gamma ) =0 $ is also valid for local
BRS transformations. This can be written using ~\ref{ds} as
\begin{eqnarray}
\int d^4x\left[ \partial _\mu (D_\nu c F^{\mu\nu}) \lambda
+\frac{1}{ \lambda } \partial _\mu ( \partial\cdot A D_\mu c)
\Lambda   - \partial _\mu (K^\mu \frac{g}{2}fcc ) \Lambda
\nonumber\right.  && \\
+\left.j^\mu <D_\mu c \Lambda >+\bar{\xi}<-\frac{g}{2} fcc
\Lambda > + <-\frac{1}{ \lambda }( \partial\cdot A+t) \Lambda >\xi
 \right] = 0 &&
 \end{eqnarray}
 This can be further written by using Eqs.~\ref{ni11} - \ref{ni13}
  as
 \begin{eqnarray}
 &&\int d^4x \frac{\partial S}{ \partial A_\mu(x)} \delta _2 A_\mu
 (x) + \frac{1}{ \lambda }\int d^4x \left\{ \partial _\mu [
 \partial ^\mu (( \partial\cdot A+t) c)]+\partial _\mu(( \partial\cdot
 A +t)D_\mu c )\right\} \Lambda
 \nonumber \\ && \mbox{\hspace{.3in}} -\int d^4x \Big\{ \partial _\mu \big[
(K^\mu
 + \partial ^\mu \zeta )
 \left.\frac{g}{2} fcc \big] \right\}\Lambda
 + j^\mu  <D_\mu c \Lambda >
 +\bar{\xi} <- \frac{g}{2} fcc \Lambda > +\nonumber \\
 &&\mbox{\hspace{2.2in}}<-\frac{1}{ \lambda }
 ( \partial\cdot A +t) \Lambda > \xi = 0
 \label{big}
 \end{eqnarray}
 We now use the equation of motion
 \begin{equation}
 < \left(\frac{ \delta S}{ \delta A_\mu} + j_\mu +fKc\right)
 \delta _2 A_\mu >=0
 \end{equation}
 Dropping $ \Lambda $ from the above expression and making some
 simplification we can write Eq.~\ref{big} as
 \begin{eqnarray}
< \partial _\mu (j_\mu c) + \frac{1}{ \lambda } \partial _\mu [(
 \partial\cdot A+t) gfc A_\mu ] + \frac{1}{ \lambda } \partial _\mu
 \left[ \partial _\mu [ ( \partial\cdot A +t)]c - (\partial\cdot A +t)\partial
_\mu
 c \right]- \nonumber && \\
  \partial ^\mu \left\{ (K^\mu + \partial ^\mu \zeta
 )\frac{g}{2} fcc \right\}> + j_\mu <D_\mu c>  + \bar{\xi}<-\frac{g}{2}
 fcc > - <-\frac{1}{ \lambda }( \partial\cdot A +t) > \xi =0 \nonumber && \\
 \label{big1}
 \end{eqnarray}

Now we note that
\begin{eqnarray}
\frac{1}{ \lambda }\left [ \partial _\mu ( \partial\cdot A +t )c -
(\partial\cdot A +t) \partial
_\mu c \right ] - \frac{g}{2} \partial ^\mu \zeta fcc + \frac{1}{ \lambda }
g ( \partial\cdot A +t) f cA_\mu \nonumber && \\=
-\frac{1}{ \lambda }\left( \partial _\mu ( \partial\cdot A +t)c\right ) +
\delta _{BRS}\left (
2 \partial _\mu \zeta c + gf \zeta cA_\mu \right )
\label{brs}
\end{eqnarray}
[ For any function $f$, change under global BRS $\equiv \delta _{BRS} f \Lambda
$
]. Then we can write the local BRS WT identities as
\begin{eqnarray}
j_\mu <D_\mu c> + \bar{\xi}< -\frac{g}{2} fcc> + \frac{1}{ \lambda } ( \partial
.A +t)> \xi\mbox{\hspace{.5in}} \nonumber && \\
= - \partial _\mu <j_\mu c + \frac{g}{2}K^\mu fcc > - \partial _\mu <
\delta _{BRS} O^\mu > + \frac{1}{ \lambda } \partial ^2 <( \partial .A +t)c>
\label{o}
\end{eqnarray}
with
\begin{equation}
O^\mu = 2 \partial ^\mu \zeta c + f \zeta c A^\mu
\end{equation}

We note that, if we integrate the equation ~\ref{o} the right hand side
being a total divergence contributing nothing and we recover the usual global
BRS WT identity. The first two terms on the right hand side will however
contribute to the ``first moment'' of the equation obtained by multiplying
by $ \epsilon \cdot x $ and integrating over $d^4x$ [ $\epsilon _\mu $
is an arbitrary constant four-vector]; while the last term will only
contribute to the  ``second moment''. We shall be particularly
interested in the ``first moment'' equation.
\begin{eqnarray}
\int d^4x \epsilon \cdot x < j^\mu D_\mu c + \bar{\xi}(-\frac{g}{2} fcc)
+\frac{1}{ \lambda } ( \partial \cdot A +t) \xi > \nonumber && \\
= \int d^4x \epsilon _\mu < j^\mu c + \frac{g}{2} K^\mu fcc + \delta _{BRS}
O^\mu > \label{mom}
\end{eqnarray}
This will be exploited in the next section to relate the above equation
to the broken OSp invariance.

\section{Connection with Broken OSp Symmetry}
In this section, we shall make contact with the superspace formulation
of Ref.\cite{osp}. We shall then see how the equation ~\ref{mom} from
local BRS is equivalent to the broken OSp symmetry of Eq. ~\ref{ww}.

In order to do this, we note the correspondence between the sources
in $W$ of Eq. ~\ref{ww1} and $ \bar{W} $, superspace generating
functional, as exhibited
in Eq. ~\ref{t275}. This requires $ j^\mu \rightarrow K^\mu_{, \theta }
, \bar{\xi} \rightarrow K^5 _{, \theta }, \xi \rightarrow -t_{, \theta }
, \kappa \rightarrow K , l \rightarrow K^5 .$ Then Eq.~\ref{mom}
reads
\begin{eqnarray}
\int d^4x \epsilon \cdot x < K^\mu_{, \theta } D_\mu c + K^5_{ , \theta }
( -\frac{1}{2}g fcc ) -\frac{1}{ \lambda }( \partial \cdot A +t ) t_{, \theta }
> \nonumber && \\
= \int d^4x \epsilon _\mu <K^\mu_{, \theta }c + \frac{1}{2} g K^\mu fcc +
\delta
_{BRS} O^\mu > \label{ffo}
\end{eqnarray}
To convert this into a equation for $ \bar{W} $ we note the correspondence
from Eq. ~\ref{t275}. In particular note
\begin{eqnarray}
\frac{ \delta \bar{W} }{ \delta K^\mu} = \prod \delta (K^4)\prod K^4_{, \theta
}
\frac{ \delta W}{ \delta K^\mu} &=& \prod \delta ( K^4) \prod K^4_{, \theta }
< D_\mu c > iW \nonumber \\
&=&<D_\mu c> i\bar{W} \equiv i\ll D_\mu c\gg
\label{ft}
\end{eqnarray}
and
\begin{eqnarray}
\frac{ \delta \bar{W} }{ \delta K^5_{, \theta }} = \prod \delta (K^4)\prod
K^4_{, \theta }
\frac{ \delta W}{ \delta \bar{\xi}} &=& \prod \delta ( K^4) \prod K^4_{, \theta
}
< c  > \nonumber \\
&=& <c>i\bar{W} \equiv i\ll c\gg
\label{ft1}
\end{eqnarray}
and analogous equation for $\frac{ \delta \bar{W} }{ \delta K^5}, \frac{ \delta
\bar{W} }{ \delta t} $. We multiply Eq. ~\ref{ffo} by a $ \prod \delta (K^4)
\prod
K^4_{, \theta }W$ [ $a$ =  an anticommuting constant.] and use Eqs. ~\ref{ft}
and
{}~\ref{ft1} in it to obtain
\begin{eqnarray}
\int d^4 x \epsilon \cdot x a \left[ K^\mu_{, \theta } \frac{ \delta \bar{W}
}{ \delta K^\mu}+ K^5_{, \theta } \frac{ \delta \bar{W} }{ \delta K^5}
 +t_{, \theta }\frac{ \delta \bar{W} }{ \delta t}\right] \nonumber\mbox{
 \hspace{1.2in}}
 && \\ - \int d^4x \left[ \epsilon _\mu a K^\mu_{, \theta }\frac{ \delta
 \bar{W} }{ \delta K^5_{, \theta }}- \epsilon _\mu a K^\mu \frac{ \delta
 \bar{W} }{ \delta K^5}\right] = i\int \delta _{BRS} \ll \epsilon _\mu
 a O^\mu \gg d^4x
 \label{ff}
 \end{eqnarray}
 On account of the fact that $ \bar{W} $ contains $ \prod K^4_{, \theta };
 \, K^4_{, \theta } \frac{ \delta \bar{W} }{ \delta K^4} =0 $ and hence this
term
 can be added to the first square bracket on the left hand side of Eq.
{}~\ref{ff}
 . Also for similar reasons, $\epsilon _\mu a K^4 \frac{ \delta W}{ \delta
K^\mu}
 + \epsilon ^\mu a K^4_{, \theta}\frac{ \delta W}{ \delta K^\mu} $ can be added
to the
 second square bracket. Then the left hand side of Eq.~\ref{ff} is just the
change in
 $ \bar{W} [ \bar{K} ,t] $ under the infinitesimal transformations:
 \begin{eqnarray}
 \delta K^\mu (\bar{x}) &=& \epsilon \cdot x a K^\mu_{, \theta } (\bar{x}) +
 \epsilon _\mu a K^4 (\bar{x}) \nonumber \\
 \delta K^5 (\bar{x}) &=& \epsilon \cdot x a K^5_{, \theta }(\bar{x}) +
 \epsilon _\mu a K^\mu (\bar{x}) \nonumber \\
 \delta K^4 &=& 0, \,\,\,\, \, \delta t = \epsilon \cdot x a t_{, \theta }
 \label{ff1}
 \end{eqnarray}
 and the ones they, in particular, contain
 \begin{eqnarray}
 \delta K^\mu_{, \theta } &=& - \epsilon _\mu  a K^4_{, \theta
}\mbox{\hspace{1in}}
 \delta K^4_{, \theta } =0 \nonumber \\
 \delta K^5_{, \theta } &=& - \epsilon _\mu a K^\mu _{, \theta
}\mbox{\hspace{1in}}
 \delta t_{, \theta } = 0
 \label{fs}
 \end{eqnarray}
 But these are just the superspace rotation transformations for an
anticommuting
 4-vector $ K^\mu (\bar{x}) $ and a scalar $t (\bar{x})$

\begin{eqnarray}
K^{\prime \mu}(\bar{x}^\prime )&=& K^{\prime \mu}(x+ \epsilon a
\lambda , \lambda , \theta - \epsilon \cdot x a) = K^\mu (\bar{x})
+ \epsilon _\mu a K^4 (\bar{x}) \nonumber \\
K^{\prime 5} (\bar{x}^\prime ) &=& K^5 (\bar{x}) + \epsilon _\mu a
K^\mu (\bar{x}) \label{fs1}
\end{eqnarray}
Where we have ignored the translation in $ x $ contained in
Eq.~\ref{fs1} because it leads to no change in $ \bar{W} $ on
account of its translational invariance. Thus Eq.~\ref{ff} just
reads
\begin{equation}
\bar{W} [ \bar{K}^\prime ,t^ \prime  ] = \bar{W} [K ,t] + i\int
\delta _{BRS} \ll \epsilon _\mu a O^\mu \gg d^4x
\label{fe}
\end{equation}
It can be easily shown that in the superspace formulation of
Ref.~\cite{osp}, the last term in Eq.~\ref{fe} is equal to the last
term in Eq.~\ref{ww}, thus proving the broken $ OSp(3,1|2) $
identities of Eq.~\ref{ww} directly via ``local BRS''.

\section{Conclusions and Observations}
  In this section, we summerize the conclusions of this work and make a
  number of further observations that could lead to new results.

  To summerize, we have introduced a straightforward local generalization
  of BRS transformation, which we call ``local BRS ". This is not an
  exact symmetry of the Yang-Mills effective action. However, it leads to a
  WT identity which (i) contains the usual BRS WT identity as a special case.
  (ii) contains an extra operator. The presence of an extra operator at first
sight
  deters one from being able to extract new consequences with case. But as we
shall
  observe later, the renormalization properties of this operator are
particularly
  simple and this may enable one to derive new consequences.

  In this work, we have however focussed on another formal aspect, the
connection
  between this broken ``local BRS " symmetry and broken OSp symmetry discussed
  recently in the context of a superspace formulation of gauge
theories~\cite{osp}. We have shown
   that the ``local BRS " WT identity is precisely equivalent to the broken
$OSp(3,1|2)$
   WT identity obtained earlier using the broken symmetry of superspace
formulation
   under superrotations that mix $x_\mu$ with $\lambda $ and $\theta $. We thus
give a
   backward derivation of approximate $OSp(3,1|2)$ symmetry of Yang-Mills
theories.

   Finally we make a number of observations:

   (1) It is well known from the renormalization transformation
   properties of gauge theories ~\cite{le} that the left hand side of
   \ref{mom} suffers an {\it overall} renormalization $({Z\tilde{Z}})^
   {-\frac{1}{2}}$. Hence multiplying the Eq.~\ref{mom} by $({Z\tilde{
   Z}})^{\frac{1}{2}}$, we obtain, in particular
   \begin{eqnarray}
   \mbox{finite} &=& \tilde{Z} \int \left\{j^{\mu R}<c>^R + K^{\mu
   R}<\frac{1}{2} gfcc >^R +<\delta_{ BRS } O^\mu>\right\}d^4x \nonumber
   \\
   &=& \tilde{Z}\int d^4x \left\{-\frac{\delta\Gamma}{\delta A_\mu^R}c^R
   -K^{\mu R}\frac{\delta\Gamma}{\delta l^R} +<\delta_{ BRS
   }O^\mu>\right\}
   \end{eqnarray}
   This immediately gives the renormalization properties of $<\delta_{
   BRS} O^\mu>$ : At $K=0$, it receives counter terms proportional to
   itself and to $\frac{ \delta S}{\delta A_\mu} c$ both dependent on
   $\tilde{Z}$, a ``known" renormalization constant. This observation
   should help in finding possible applications of ~\ref{mom}

   (2) The operator $O^\mu$ can, in fact, be further simplified. We note
   \begin{eqnarray}
   \delta_{BRS} O_\mu &=& \delta_{BRS}\left\{2\partial_\mu\zeta c + f\zeta c
A_\mu\right\}\nonumber \\
  &=&\delta_{BRS}\left(\partial_\mu\zeta c +\frac{\partial
.A}{\lambda}A_\mu\right) -
   \frac{1}{\lambda}\partial .D c A_\mu +\partial_\mu \delta_{BRS}(\bar{c}c)
   \label{bpm1}
   \end{eqnarray}
   When \ref{bpm1} is substituted in \ref{mom}, the last term in \ref{bpm1}
   does not contribute being a total derivative; the second to last term can be
   simplified using equation of motion of $\zeta $. The net result is
   \begin{eqnarray}
   \int d^4x \epsilon .x <j^\mu D_\mu c +\bar{\xi}(-\frac{g}{2}
fcc)+\frac{1}{\lambda}
   (\partial . A +t)\xi> = \nonumber \\
   \int d^4x \epsilon_\mu <j^\mu c +\frac{1}{2}K^\mu g fcc
+\frac{1}{\lambda}\xi A_\mu>
   +\int d^4x \epsilon_\mu <\delta_{BRS}\tilde{O}^\mu>
   \end{eqnarray}
   Where $\tilde{O}^\mu $ is the simpler looking operator
   \begin{equation}
   \tilde{O}^\mu = \partial^\mu\zeta c +\frac{\partial .A}{\lambda}A_\mu
   \end{equation}
   [ We kept $O^\mu $ in \ref{mom} as it was the relevant one to OSp WT
identity].

   (3) In view of the fact that $ x^\mu \sim i\frac{\partial}{\partial p^\mu}
$, the
   local BRS WT identity of Eq.~\ref{mom} may be useful in studying momentum
variation
   of Green's functions.


\newpage

 \begin{references}
 \bibitem{le} See e.g. B. W. Lee Phys. Lett. {\bf 46 B} 214 (1974) and
 references therein.
 \bibitem{zinn} J. Zinn-Justin Lectures at "International Summer
 Institute for Theoretical physics" Bonn 1974.
 \bibitem{lee} See, for example, B. W.Lee, in {\em{Methods in Field Theory,
 Les Houches,France,1975}} edited by R. Balian and J. Zinn-Justin
 (North-Holland, Amsterdam .)
 \bibitem{osp} S. D. Joglekar and B. P. Mandal, Phys. Rev. D{\bf 49}, 5382
 (1994).
 \bibitem{sdj} S. D. Joglekar (unpublished) ; Phys. Rev. D {\bf 43}, 1307
 (1991);{\bf 48}, 1878(E) (1993).

\end {references}
\end{document}